# ASTRONOMY AND THE FIFTH DIMENSION

Paul S. Wesson

Department of Physics and Astronomy,
University of Waterloo, Waterloo, Ontario N2L 3G1, Canada

Abstract: Astronomy is a precise and relatively simple science because objects accelerate in a gravitational field at the same rate, irrespective of their composition. Galileo knew this, and Einstein took it as the basis for general relativity. Surprisingly, it is also a consequence of new theories that use a fifth dimension.

Email: psw.papers@yahoo.ca

# ASTRONOMY AND THE FIFTH DIMENSION

Imagine, if you will, a street entertainer juggling an apple, an orange and a banana. While the juggler may not be thinking about Einstein, it is because of the latter's Equivalence Principle that the objects do not end up on the ground in a fruity mess. The juggler is probably not thinking either about the possible existence of an extra dimension in addition to the space and time of general relativity; but recent work indicates that Einstein's Equivalence Principle follows naturally from 5D relativity.

Einstein's Equivalence Principle (EEP) says in the simplest form that objects in the Earth's gravitational field fall at the same rate, irrespective of their composition. This refers not only to chemical composition, but also to isotopic composition, including contributions to the measured mass from electromagnetic and nuclear forces. Without the EEP, it would be difficult to estimate the orbit of any object, either on the Earth or in space. It is the EEP which guarantees the juggler's art, and it is also the reason why astronomy is such an ancient and exact branch of science.

Historically, physicists have distinguished between three different types of mass. Active gravitational mass is the quantity which is responsible for the force, passive gravitational mass is the one which responds to the force, and inertial mass is the thing which resists acceleration and also measures the energy content of an object (given by Einstein's famous formula). The first two types of mass can readily be shown to be proportional to each other by reciprocity arguments, so it is the proportionality of gravitational mass $m_g$ to inertial mass $m_i$ which figures in the EEP.

To see why the EEP is important, consider the Kepler problem of the Earth in orbit around the Sun. Using large and small letters for the masses, the gravitational force between the Sun and the Earth is $GM_g m_g / r^2$ where $G$ is Newton's constant and $r$ is the separation. This attraction is countered by the centrifugal or inertial force $m_i v^2 / r$ where $v$ is the Earth's orbital velocity. As an equation we write

$$\frac{GM_g m_g}{r^2} = \frac{m_i v^2}{r} \quad , \tag{1}$$

and we *cancel* the $m_g$ on the left-hand side with the $m_i$ on the right-hand side. This is allowable because the EEP says that the two masses are proportional to each other (and usually set equal). The result is a Kepler orbit for the Earth, with $v = \sqrt{GM/r}$, as every student knows.

It need not be so, however. Some modern theories of gravitation, which go beyond that of Einstein, predict extra accelerations that can be tested using astronomical observations. The bases of these theories are various, but several of them include a scalar field which (like ordinary gravity and electromagnetism) acts over large distances, and can in principle modify the dynamics of objects in the solar system and beyond. Particular attention has been paid to the orbit of the Moon, whose effective mass depends on its binding energy; and on the trajectory of the *Pioneer* spacecraft, which appears to show an anomalous acceleration. However, when the observational uncertainties are taken into account, there is no compelling evidence for a departure from general relativity, so as far as present data go, Einstein's Equivalence Principle holds. That it does so to reasonable accuracy has been known for centuries, and at least since the time when Galileo (supposedly) dropped balls from the Leaning Tower of Pisa.

Indeed, the EEP is often taken for granted. However, some great thinkers have sought a deeper rationale for the proportionality of gravitational and inertial mass, and some clever experimenters have verified the fact to great accuracy (of order 1 part in $10^{12}$). Einstein was, of course, motivated by Equivalence to formulate the general theory of relativity; and indirect support for the EEP comes from the numerous tests of that theory. Recently, general relativity was confirmed by measuring the precession rates of a set of super-cooled gyroscopes aboard a drag-free satellite in Earth orbit[1]. Another reason for carrying out the experiment was to look for possible departures from Einstein's theory. In that regard, it is widely believed that the best route to a unification of gravity with the interactions of particles is by a theory like general relativity, but with more dimensions than the familiar 4 of spacetime. Five-dimensional relativity is the basic extension, and there is in fact a small difference in the precession rate of a gyroscope in a gravitational field, depending on whether the world has 4 or 5 dimensions[2]. This particular effect proved too small to detect by experiment. However, more work on 5D relativity has newly revealed implications for the status of the EEP[3]. It appears, in fact, that the EEP may be a direct consequence of the existence of an extra dimension.

To see why, let us add to the 4 standard coordinates of space and time an extra one, say $\ell$. To calculate dynamical effects, we need to write down an expression for the square of the 'distance' between two nearby points (given by the interval $dS^2$) and find the 'shortest' path between the two points (given by the extremum of $S$). This procedure is analogous to using Pythagoras' theorem to find the shortest path between two points in ordinary 3D space. We do not – to start with – know how to visualize the 'shape' of a 5D world. However, other work on 5D relativity has led to a particularly simple form which is called canonical 5D space[4]. For this, the shape in 'cross-section' resembles an ordinary circle. A circle drawn on a flat surface like this page is best described in terms of the radius $r$ and the angle $\theta$ which sweeps around counter-clockwise from a given starting point. The square of the distance between two nearby points is then $d\sigma^2 = dr^2 + r^2 d\theta^2$. In this formula, we can replace $r$ by $\ell$, and replace the increment of angle $d\theta$ by the ratio of two lengths $ds/L$, where $ds$ is the interval of Einstein's 4D general relativity and $L$ is a constant length whose meaning will soon be made clear. It is also instructive to rearrange the terms and swap a sign to indicate that the new $\ell$ is physically like a measure for space rather than a measure for time. The result is the 5D interval

$$dS^2 = (\ell/L)^2 ds^2 - d\ell^2 \quad . \tag{2}$$

This defines 'distance' in 5D, and involves a term like that in 4D ($ds^2$) and a term to do with the extra dimension ($d\ell^2$), though the two parts are interdependent (and in the general case $ds^2$ may conceal an internal dependency on $\ell$). However, while it may look strange, the equation just given is still basically that of a circle.

Paths in the 5D world described by (2) can be obtained by following a standard procedure[4]. Two remarkable things emerge. (a) The paths of all particles, even massive ones, can be described by (2) with $dS^2 = 0$. This means that a particle with a finite mass $m$ follows the same kind of *null* path in 5D that a massless photon follows in 4D (where $ds^2 = 0$). A corollary of this is that *all* particles are in some kind of causal contact in 5D, since $dS^2 = 0$ now takes the place of causal contact defined by the exchange of light signals with $ds^2 \geq 0$ in 4D. (b) The paths of particles are generally affected by an extra force which is associated with movement through the extra dimension. This does not upset the law of conservation of (linear) momentum, but the mass $m$ of a test particle will in general vary now along its path. This happens at a slow rate governed by the length $L$ in (2), which turns out to be related to the cosmological constant by $\Lambda = 3/L^2$. Since $\Lambda$ measures the energy density of apparently empty space, there is also a connection to the physics of the vacuum. The motion of a particle in the extra dimension is reversible, like the motion in ordinary 3D space in the absence of friction. However, via the conservation of momentum, there is now a relation between the rate at which a particle of mass $m$ varies with proper time $s$ and the rate of change of the extra coordinate $\ell$. Technically, $(1/m)(dm/ds)$ is proportional to $(1/\ell)(d\ell/ds)$, where however the motion in the extra dimension is *reversible* so $d\ell/ds$ can be positive or negative. The result of this is that there are two choices for how the mass is related to the extra coordinate. These choices are given by the proportionalities

$$\ell \sim m \quad \text{or} \quad \ell \sim \frac{1}{m} \quad , \tag{3}$$

depending on the direction of motion in the fifth dimension. This is intriguing. It is in fact the analog for massive particles in gravitational theory of the situation for electrically-charged particles in quantum theory, where following Stueckelberg and Feynman a positron may be regarded as a time-reversed electron.

The preceding is based on canonical space with interval (2), which is typical of the approach to 5D relativity known as Space-Time-Matter theory. The rationale for this theory is that mass and matter are properties of 4D spacetime which owe their existence to the fifth dimension, something which follows from Campbell's embedding theorem of differential geometry. An alternative approach to 5D relativity, known as Membrane theory, typically employs a different kind of space with an interval that is warped by the extra dimension. The rationale for this theory is that the masses of particles are controlled by a singular surface or membrane, about which matter is concentrated, thereby defining spacetime. In regard to the comments of the preceding paragraphs, it should be mentioned that the two things noted before – namely null paths and an extra force – also exist in the second theory[5]. Indeed, the mathematical structure of the two theories is similar[6,7], and their conceptual bases overlap somewhat[8-10]. That said, we continue the discussion using the first approach, and focus attention on the relations (3) which show how to define the mass of a test particle.

The relations (3) are in fact the geometrical representations of how mass appears in gravitational theory and quantum theory. These branches of physics are characterized by their constants: Newton's constant $G$ and Planck's constant $h$, where the speed of light $c$ is shared by both branches in their relativistic formulations. The proportionalities (3) correspond to how gravitational and inertial mass are measured by

$$\ell_g = \frac{Gm}{c^2} \quad \text{or} \quad \ell_i = \frac{h}{mc} \quad . \tag{4}$$

These are of course the Schwarzschild radius and the Compton wavelength, as they appear in general relativity and quantum mechanics.

The implications of the preceding account are far reaching. For example, it opens the prospect of better understanding the notorious cosmological 'constant' problem, which consists basically in the discrepancy between the sizes of this parameter as derived from astrophysics and particle physics. This problem is presently the subject of intense debate. But it is apparent that in a 5D world described by canonical space (2), a connection can be made between the $\Lambda = 3/L^2$ noted above and the product of the lengths in (4), which is $Gh/c^3$ or the square of the Planck length. There are also implications of the 5D theory which are more qualitative in nature but can be tested by refining cosmological data. Notably, there are solutions of the field equations of the theory which, while they are curved in 4D, are *flat* in 5D. This means that while there may be a big bang in 4D, the universe is smooth and free of singularities in 5D. This sounds odd, but it can be demonstrated by having a computer draw the relevant plots, where the big bang appears as a point embedded in a flat background[10]. You can appreciate the same thing by taking a sheet of paper and rolling it into the shape of a cone. The sheet is intrinsically flat, but you have created the point at the apex of the cone (the 'big bang') by changing the shape, or in other words by *how* you describe the surface.

The implications for the Einstein Equivalence Principle are more straightforward. Gravitational mass and inertial mass are two aspects of the same thing, which presents itself in the ways given by (4). These, however, merely represent two ways of measuring motion in the fifth dimension. The motion is actually reversible, and there is no meaningful difference between going in the gravitational 'direction' and the inertial 'direction'. History also plays a part in our habit of representing mass in two different ways, because some thought shows that the constants *G* and *h* have to do with the ways in which gravitational physics and quantum physics have developed. The disposability of these constants and *c* is revealed by the fact that the values of all three can be consistently set to 1 by a suitable choice of units, a ploy used by researchers every day. It is clear that there is no fundamental difference between gravitational and inertial mass, so Einstein's Equivalence Principle is safe. The dedicated theorist, labouring over his arcane equations, might even suggest that the observational astronomer peering through his telescope owes something to the fifth dimension.

Acknowledgements

This article is based on previous work with other members of the Space-Time-Matter group (http://5Dstm.org).  For those interested in further reading, ref. 7 below is a mathematical review, while refs. 8-10 are non-technical accounts.